\documentclass[sigconf,screen,nonacm]{acmart}
\AtBeginDocument{%
	}
\usepackage{microtype}    
\usepackage{xcolor}       
\usepackage{comment}      
\usepackage{pifont}       
\usepackage{multicol}     

\usepackage{graphicx}     
\graphicspath{{../image/}{image/}}
\usepackage{pgfplots}     
\usepackage{subcaption}   
\usepackage{caption}      

\usepackage{booktabs}     
\usepackage{multirow}     
\usepackage{array}        
\usepackage{threeparttable} 

\usepackage{amsmath}     
  
\usepackage{amssymb}      
\usepackage{mathtools}    
\usepackage{amsthm}       
\usepackage{siunitx}       

\usepackage{tcolorbox}    
\tcbuselibrary{breakable, skins} 

\usepackage{cleveref}

\renewcommand\footnotetextcopyrightpermission[1]{}
\renewcommand{\shortauthors}{Anonymous Author(s)}

\AtBeginDocument{%
  }

\begin{document}

\title{Read Between the Stickers: Sentiment-Prior Reasoning with Learnable Verbalized Rules for Multimodal Chat Analysis} 

\author{Zixiang Ni}
\email{nzx@stu.xjtu.edu.cn}
\affiliation{%
  \institution{School of Software Engineering, Xi'an Jiaotong University}
  \city{Xi'an}
  \country{China}
}

\author{Yifei Xu}
\authornote{Corresponding authors.}
\email{belonxu_1@xjtu.edu.cn}
\affiliation{%
  \institution{School of Software Engineering, Xi'an Jiaotong University}
  \city{Xi'an}
  \country{China}
}

\author{Haowen Yang}
\email{22009200573@stu.xidian.edu.cn}
\affiliation{%
  \institution{College of Computer Science and Technology, Xi'an University of Electronic Science and Technology}
  \city{Xi'an}
  \country{China}
}
	
\author{Yang Liu}
\email{liuyang19971122@gmail.com}
\affiliation{%
  \institution{Nexus Lab, Peking University}
  \city{Beijing}
  \country{China}
}

\author{Ziyang Peng}
\email{pzytracy@stu.xjtu.edu.cn}
\affiliation{%
  \institution{School of Software Engineering, Xi'an Jiaotong University}
  \city{Xi'an}
  \country{China}
}
	
\author{Wenlong Li}
\email{wenlongli@stu.xjtu.edu.cn}
\affiliation{%
  \institution{School of Software Engineering, Xi'an Jiaotong University}
  \city{Xi'an}
  \country{China}
}
	
\author{Tingting Xin}
\email{wenlongli@stu.xjtu.edu.cn}
\affiliation{%
  \institution{Computer Science and Technology, Xi'an Jiaotong University}
  \city{Xi'an}
  \country{China}
}
	
\author{Yan Liang}
\email{liangyan514@zju.edu.cn}
\affiliation{%
  \institution{School of education, Zhejiang University}
  \city{Hangzhou}
  \country{China}
}
	
\author{Yancheng Chen}
\email{chenyancheng22@mails.ucas.ac.cn}
\affiliation{%
  \institution{Academy of Mathematics and Systems Science, Chinese Academy of
  	Sciences}
  \city{Beijing}
  \country{China}
}
	
\author{Bin Chong}
\authornotemark[1]
\email{chongbin@pku.edu.cn}
\affiliation{%
  \institution{National Engineering Laboratory for Big Data Analysis and Applications, Peking University}
  \city{Beijing}
  \country{China}
}
	
\author{Yuan Rao}
\email{raoyuan@mail.xjtu.edu.cn}
\affiliation{%
  \institution{School of Software Engineering, Xi'an Jiaotong University}
  \city{Xi'an}
  \country{China}
}

\renewcommand{\shortauthors}{Ni et al.}

\begin{abstract}
	
Multimodal chat analysis of social media stickers (MCAS) benefits from jointly modeling text and sticker semantics, yet it is inherently challenged by the interference between sentiment and intent recognition. 
Although existing multi-task approaches achieve competitive performance, they largely ignore this inter-task interference and offer little explicit reasoning about how these two predictions are made. 
To address this issue, we propose \textbf{ExCoVer}, an \textbf{Ex}plicit \textbf{C}hain-\textbf{o}f-Thought framework with \textbf{Ver}balized rules learning that integrates sentiment-prior reasoning with learnable discrimination rules to produce explicit reasoning chains for sentiment and intent predictions.
Specifically, ExCoVer consists of two components: (1) Sentiment-Prior Chain-of-Thought (SP-CoT), which detects cross-modal sentiment conflicts and uses the dominant sentiment as a prior to mitigate inter-task interference and narrow the candidate intent space; 
and (2) Verbalized Rules Learning for Confusing Intent Discrimination (VRLCID), which treats discrimination rules as learnable parameters and optimizes them via learner, optimizer, and regularizer agents to suppress spurious correlations and distinguish confusing intents.
Extensive experiments on CSMSA and MSAIRS datasets demonstrate that ExCoVer achieves state-of-the-art performance while providing explicit reasoning chains.

\end{abstract}




\maketitle

\section{Introduction}
\label{sec:introduction}

As social media proliferates, stickers have emerged as a dominant communicative modality that combines visual and textual elements to convey richer semantics than text alone, while fulfilling diverse pragmatic functions in dialogues, such as amplifying emotions, implying latent intents, and alleviating conversational awkwardness~\cite{french2017image, shi2025impact}. 
This motivates the task of Multimodal Chat Analysis of Social Media Stickers (MCAS), which supports important applications including understanding social interactions~\cite{shi2025impact}, detecting cyberbullying~\cite{lee2021disentangling}, and optimizing human-computer interaction~\cite{hu2025emotion}.
The core challenge of MCAS lies in jointly inferring sentiments and intents from multimodal sticker content, where the two subtasks are pragmatically entangled yet exhibit asymmetric mutual constraints that hinder accurate and interpretable prediction.

\begin{figure}[t]
    \centering
    \includegraphics[width=\linewidth]{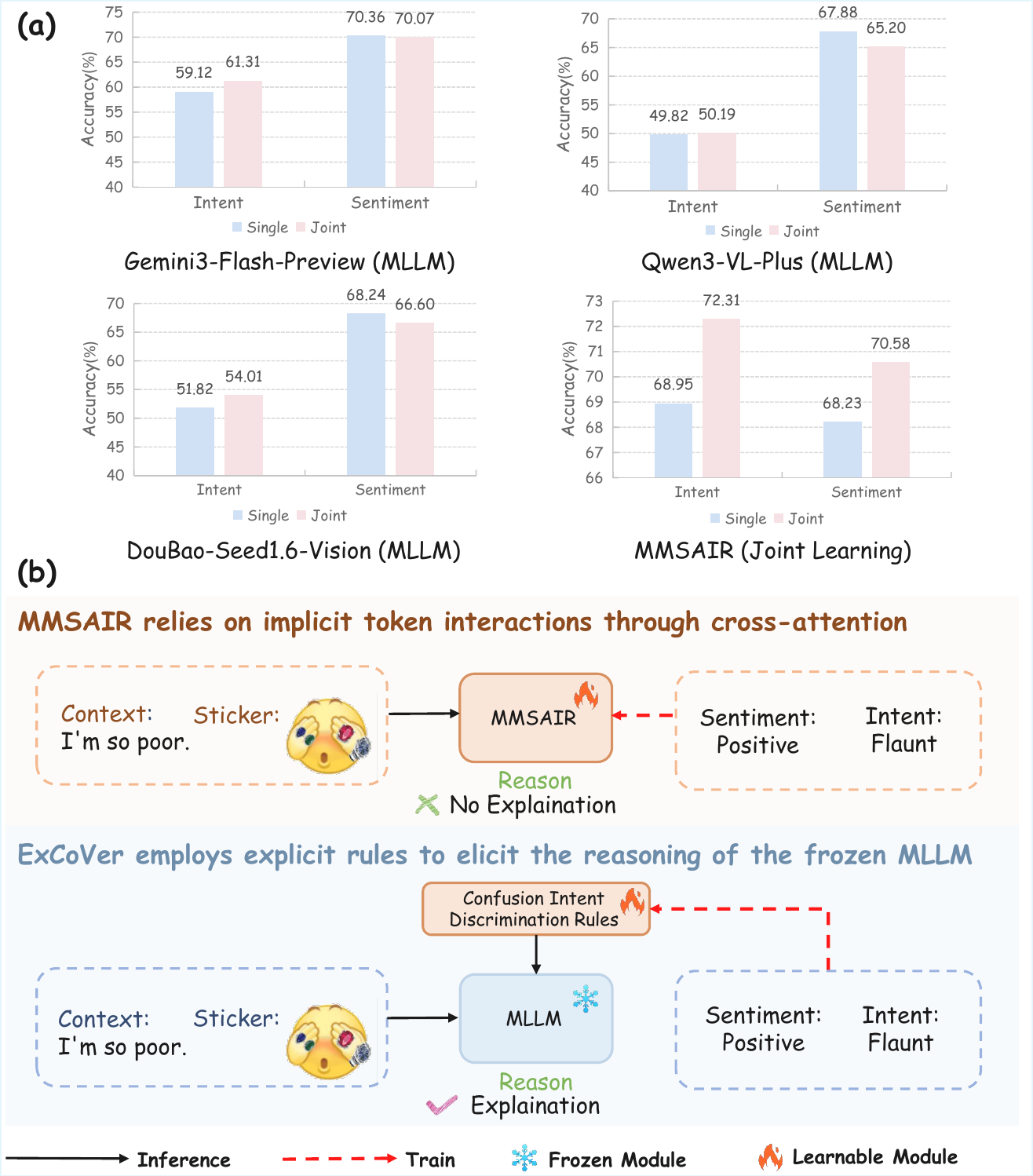}
    \caption{The illustration of our motivation. (a) Performance of MLLMs and MMSAIR under joint and single-task settings. (b) Overview of MMSAIR and our proposed ExCoVer. The brief reasoning chain of the example generated by MLLM is: \textit{``The user claims to be financially strained, yet the jewelry accessories worn by the character in the sticker achieve the effect of flaunting wealth by creating a humorous contrast''}.}
    \label{fig1:motivation}
    \Description{Motivation figure with two panels. Panel (a) compares joint-task and single-task performance of MLLMs and MMSAIR on sentiment and intent recognition. Panel (b) overviews the MMSAIR baseline and the proposed ExCoVer framework, illustrated with a sticker example about financial strain versus flaunting jewelry.}
    \vspace{-2em}
\end{figure}

To tackle MCAS, existing approaches predominantly adopt joint learning over shared multimodal encoders, assuming that sentiment and intent can mutually reinforce each other during training~\cite{liu2024emotion, shi2025impact}.
However, these methods suffer from two fundamental limitations:
(1) Adverse inter-task interference. Figure~\ref{fig1:motivation}(a) shows that jointly executing the two tasks consistently improves intent accuracy but degrades sentiment performance. This asymmetry indicates that sentiment serves as a helpful prior for intent prediction, whereas intent reasoning introduces interference that harms sentiment judgment during inference. Optimizing solely for training-time complementarity therefore overlooks such adverse dynamics at test time.
(2) Lack of explicit reasoning chains. As illustrated in Figure~\ref{fig1:motivation}(b), end-to-end joint models such as MMSAIR rely on implicit cross-attention to fuse multimodal cues, yielding predictions without explicit, human-readable reasoning chains. Consequently, it remains difficult to trace how visual and textual evidence supports the inferred sentiments or intents~\cite{weber2020integrating,arrieta2020explainable}.

Recent advances in LLMs and MLLMs provide a promising alternative by enabling explicit verbalized reasoning~\cite{caffagni2024revolution,zhang2023multimodal,guo2025deepseek}. 
Nevertheless, directly applying them to MCAS still faces two critical bottlenecks:
(1) how to leverage sentiment as an affective prior to guide intent recognition while avoiding the adverse inter-task interactions identified above;
(2) how to discriminate confusing intents under substantial semantic overlap among intent categories. Relying only on abstract category-level knowledge from pre-training, MLLMs often struggle to distinguish semantically similar intents and may produce plausible yet competing reasoning chains.

To address these challenges, we propose \textbf{ExCoVer}, an \textbf{Ex}plicit \textbf{C}hain-\textbf{o}f-Thought framework with \textbf{Ver}balized rules learning for MCAS that produces explicit reasoning chains for sentiment and intent predictions.
Specifically, ExCoVer comprises two components: (1) Sentiment-Prior Chain-of-Thought (\textbf{SP-CoT}) employs a hierarchical four-stage pipeline that detects cross-modal sentiment conflicts, derives the dominant sentiment, and uses it as a prior to guide scenario reconstruction and intent recognition, thereby mitigating adverse inter-task interference and narrowing the candidate intent space;
(2) Verbalized Rules Learning for Confusing Intent Discrimination (\textbf{VRLCID}) builds on verbalized learning~\cite{xiao2025verbalized} by formulating confusing-intent discrimination rules as learnable parameters and optimizing them through a learner--optimizer--regularizer workflow, where the regularizer filters instance-specific noise and suppresses spurious correlations to support discrimination among semantically similar intents.
Inference is performed with a frozen MLLM without additional parameter updates.
We evaluate ExCoVer on the CSMSA and MSAIRS benchmarks against representative joint-learning and MLLM baselines, where it consistently outperforms strong methods such as MSA-ITEI and MMSAIR on both sentiment and intent recognition.
Ablation studies confirm that removing SP-CoT or VRLCID substantially degrades performance, validating their complementary roles and the benefits of sentiment-prior guidance and regularized verbalized rules.
The main contributions of this work are summarized as follows:
\begin{itemize}
    \item {\bf Explainable Framework.} We propose ExCoVer for MCAS, which jointly predicts sentiment and intent with explicit, human-readable reasoning chains using a frozen MLLM, without additional parameter updates.

    \item {\bf Sentiment-Prior Reasoning.} We develop SP-CoT, a four-stage pipeline that resolves cross-modal sentiment conflicts, derives the dominant sentiment, and uses it as a prior to guide scenario reconstruction and intent recognition, thereby mitigating adverse inter-task interference and narrowing the candidate intent space.

    \item {\bf Learnable Verbalized Rules.} We introduce VRLCID, which treats confusing-intent rules as learnable parameters optimized via a learner--optimizer--regularizer workflow, where the regularizer suppresses spurious correlations to distinguish semantically similar intents.

    \item {\bf Experimental Validation.} Extensive experiments and ablation studies on the CSMSA and MSAIRS datasets demonstrate that ExCoVer achieves state-of-the-art performance on both intent and sentiment recognition while providing explicit reasoning chains, and confirm the necessity of SP-CoT and VRLCID. 
\end{itemize}

\section{Related Work}
\label{sec:related}
\subsection{Multimodal Sentiment and Intent Recognition in Social Media Stickers}

Multimodal sentiment and intent analysis has attracted increasing attention, with extensive research devoted to integrating multiple modalities (e.g., text, audio, and visual signals) to deepen the understanding of human affective states and behavioral goals~\cite{li2025tracing, fang2025emoe, zhang2022mintrec, zhang2024mintrec2, li2025cusmer, xia2025effective, hu2025adaptive, zhou2025llm}. Recent studies have increasingly emphasized uncovering the interaction between sentiment and intent in social communication. Notably, \citet{liu2024emotion} demonstrates that the strong complementarity between emotion and intent enables simultaneous improvements in both tasks through joint learning~\cite{singh2022emoint, yang2025uncertain}. 
However, existing studies predominantly focus on real-world conversational scenarios, and largely overlook social media conversations that involve stickers. As compact yet highly expressive visual symbols, stickers convey subtle semantics that are difficult to capture through text alone~\cite{french2017image}. To address this gap, \citet{ge2022towards} and \citet{shi2025impact} construct the CSMSA and MSAIRS benchmarks for sticker-based sentiment and intent analysis. Later, MSA-ITEI~\cite{shi2025msa} achieves competitive performance on both datasets by employing MLLMs to generate textual descriptions for stickers. 
Despite these advances, existing sticker-based methods largely exploit sentiment-intent complementarity during training, but do not explicitly model their asymmetric interactions during joint inference.
Moreover, they typically produce predictions through implicit multimodal fusion (e.g., cross-attention in MMSAIR) or auxiliary sticker descriptions (e.g., in MSA-ITEI), without explicit reasoning chains that trace how context and sticker cues jointly support sentiment and intent predictions.
This gap motivates SP-CoT in ExCoVer, which performs explicit sentiment-prior reasoning at inference to model asymmetric sentiment--intent interactions and trace how context and sticker cues jointly support sentiment and intent predictions.

\subsection{Verbalized Learning for MLLMs}
Verbalized learning represents an emerging training paradigm that enables MLLMs to autonomously learn linguistic rules and task-specific descriptions, which offers promising potential for addressing confusing intent discrimination problems. 
In Verbalized Machine Learning (VML)~\cite{xiao2025verbalized}, verbal expressions of task criteria are treated as learnable components that undergo iterative optimization through interactions between learner and optimizer agents. 
Recent work has extended VML beyond conventional machine learning tasks to more complex domains. 
VERA \cite{ye2025vera} generalizes VML to video anomaly detection by employing visual language models (VLMs) to instantiate learner and optimizer agents that refine question sets designed to capture anomalous patterns in video sequences. 
Similarly, Verbalized Representation Learning \cite{yang2025verbalized} applies verbalized learning to few-shot image classification. 
In this setting, VLMs distill inter-class distinctions and intra-class commonalities into verbalized features, which significantly improve accuracy on fine-grained and novel concept categorization. 
The intrinsic semantic overlap between such intent pairs causes MLLMs to absorb noise from the training data or amplify model-specific biases when learning discrimination rules. 
Critically, current verbalized learning paradigms lack a dedicated regularization mechanism to prevent the learner from establishing spurious associations between incidental, instance-specific attributes and intent categories. Without such constraints, the class discrimination rules treated as learnable parameters tend to over-specialize to the idiosyncratic features of individual training samples rather than capturing generalized task logic.
This gap motivates VRLCID in ExCoVer, which extends verbalized learning with a regularizer agent to filter instance-specific noise and suppress spurious correlations when learning confusing-intent discrimination rules.

\begin{figure*}[t]
    \centering
    \includegraphics[width=\textwidth]{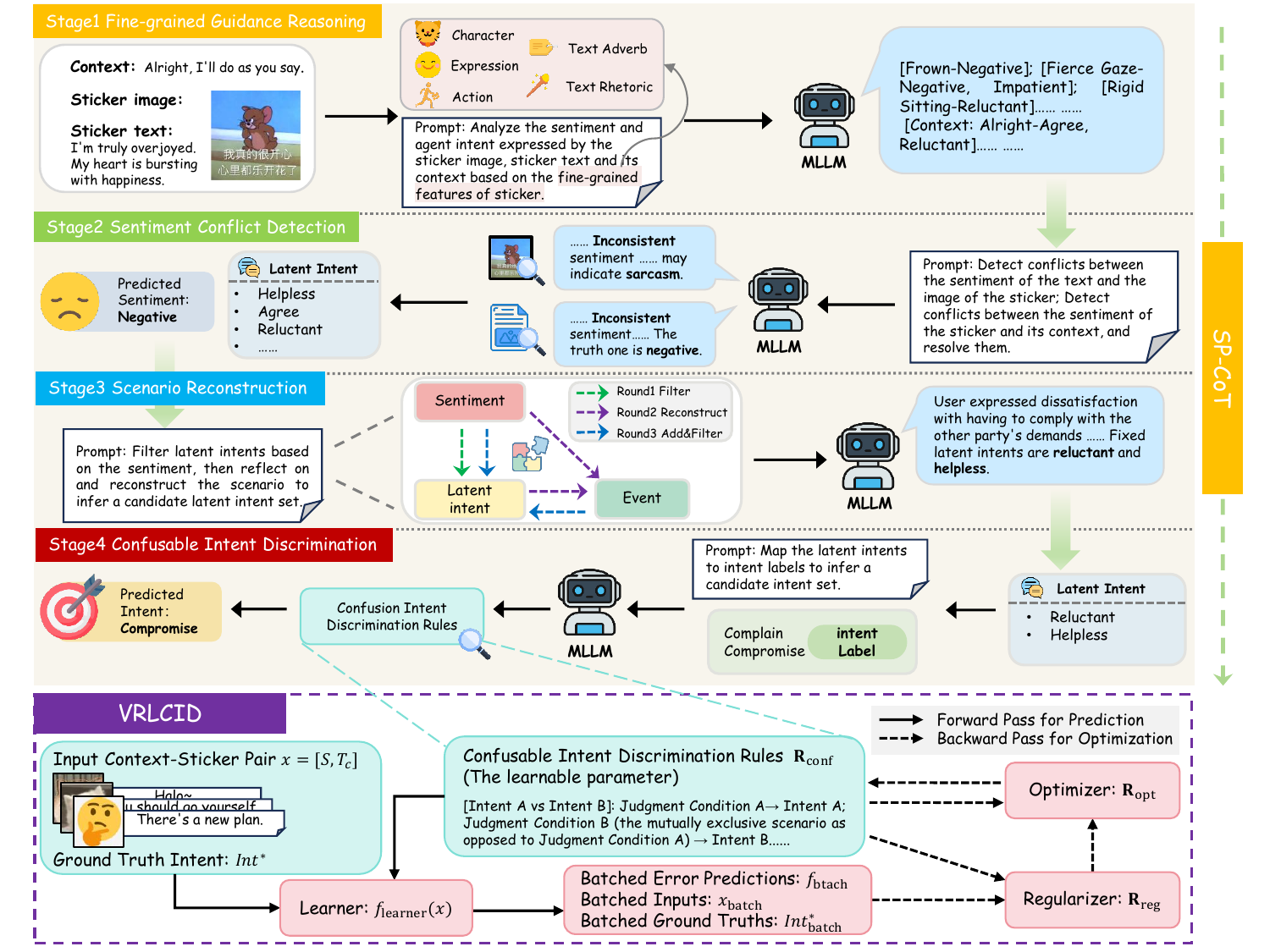}
    \caption{An overview of our ExCoVer. The frozen MLLM combines stickers and context to reason the sentiment and candidate intent set by SP-CoT. It then distinguishes the confusing intents within the candidate intent set based on the confusing intent discrimination rules learned on VRLCID, and further determines the predicted intent.}
    \label{fig3:flowchart}
\end{figure*}

\section{Methodology}
\label{sec:method}
Our proposed ExCoVer comprises two core components: a Sentiment-Prior Chain-of-Thought (SP-CoT) for reasoning, and a Verbalized Rules Learning module for Confusing Intent Discrimination (VRLCID) for learning confusing intent discrimination rules. As illustrated in Figure \ref{fig3:flowchart}, SP-CoT adopts a hierarchical reasoning strategy that first addresses sentiment recognition, then utilizes the identified sentiment as an affective prior to guide intent recognition. VRLCID enables MLLMs to explicitly model confusing intent discrimination rules through verbalized rules learning. A sticker is represented as $S = [I_s, T_s]$, where $I_s$ and $T_s$ denote its image and text components, respectively. Given a context $T_c$ and a sticker $S$ with the SP-CoT prompt template $\theta_\text{SP-CoT}$ and confusing intent discrimination rules $\textbf{R}_{\text{conf}}$ learned in VRLCID, the objective is to jointly predict intent $\hat{Int}$ and sentiment $\hat{Sen}$ based on the MLLM $\Psi_{M}$:
\begin{align}
(\hat{Sen}, \hat{Int})&= \Psi_{M}\left(S, T_c \mid \theta_\text{SP-CoT},\textbf{R}_{\text{conf}} \right)
\end{align}
\subsection{Sentiment-Prior Chain-of-Thought (SP-CoT)}
\label{sec:cot}
Inspired by the cognitive linguistic theory of reverse appraisal~\cite{de2012reverse}, which posits that inferring the cognitive appraisal underlying emotional expressions facilitates intent understanding, we propose SP-CoT to utilize sentiment as a guiding prior for intent recognition. As depicted in Figure \ref{fig3:flowchart}, SP-CoT comprises four sequential stages: Fine-grained Guidance Reasoning, Sentiment Conflict Detection, Scenario Reconstruction, and Confusing Intent Discrimination.
Specifically, we first leverage the pre-trained knowledge of MLLMs to extract fine-grained features from stickers and context to identify sentiment and latent intent. Conflict detection is then employed to resolve cross-modal sentiment inconsistencies to determine the dominant sentiment. Subsequently, we filter latent intents incompatible with sentiment and prioritize those derived from context to construct the candidate latent intent set for scenario reconstruction. The final intent is inferred by first mapping candidate latent intents to predefined labels, then applying the confusing intent discrimination rules obtained through verbalized rules learning to distinguish confusing intents and predict the final intent.

\textbf{Fine-grained Guidance Reasoning.}
To comprehensively capture critical sentiment or intent information embedded in stickers and context, the MLLM is first prompted to extract fine-grained tokens from context $T_c$ and sticker $S$. It analyzes the image modality across dimensions of character, facial expression, and action, and performs token-level analysis of the textual modality with emphasis on subtle elements such as adverbs and rhetorical devices. 
Based on this analysis, the MLLM produces initial estimates of sentiment and latent intent sets. In detail, the latent intent sets $Int^{lat}_c$ and $Int^{lat}_s$ derived from $T_c$ and $S$ encompass but are not limited to predefined intent labels and can be empty, thereby preventing the model from being constrained to forcibly associate every fine-grained token with a predefined intent label. The sentiments obtained from $T_c$, $I_s$, and $T_s$ are determined as the sentiments $Sen_c$, $Sen_{is}$, and $Sen_{ts}$, respectively.

\begin{figure}[htbp] 
    \centering 
    \includegraphics[width=0.92\linewidth]{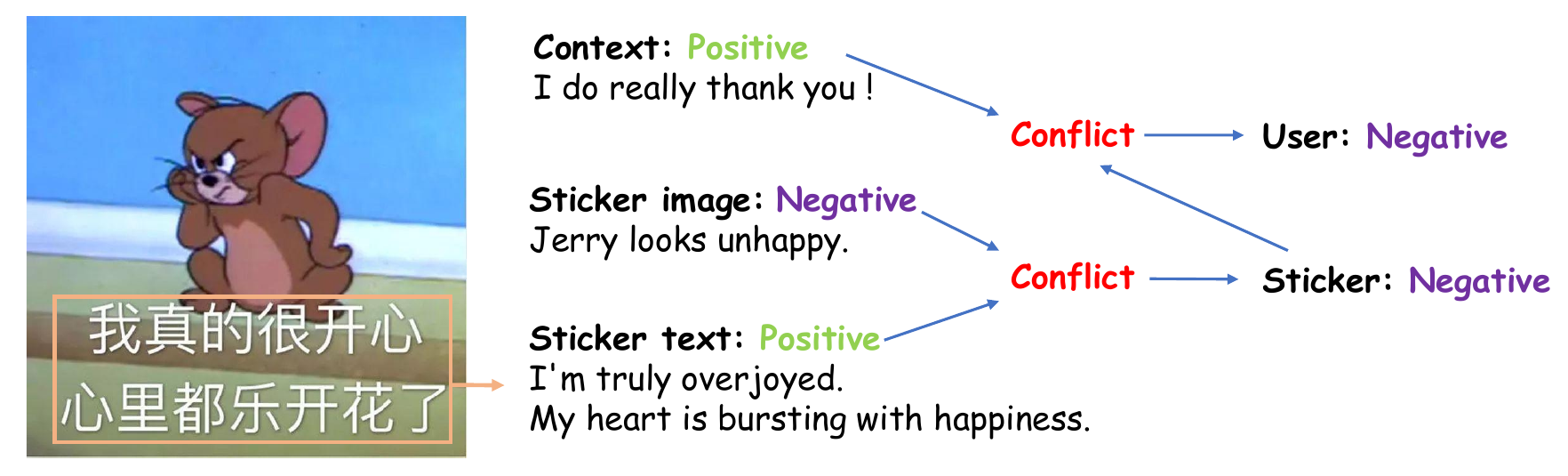} 
    \caption{The illustration of multimodal sentiment conflict in a sticker.} 
    \label{fig:conflict} 
\end{figure}

\textbf{Sentiment Conflict Detection.}
Since sentiment can serve as a powerful prior to facilitate intent recognition, we design sentiment conflict detection to effectively infer sentiment. Sentiment conflicts typically manifest at two hierarchical levels: inter-modality conflicts between context and stickers, as well as intra-sticker conflicts between the sticker image and sticker text. As illustrated in Figure \ref{fig:conflict}, we examine a sticker containing the text \textit{``I'm truly overjoyed. My heart is bursting with happiness.''} that conveys a positive sentiment. However, Jerry's irritated body language reveals ironic negative sentiment. Similarly, the context \textit{``I do really thank you!''} typically expresses positive sentiment, but it becomes evident that the user may be dissatisfied with the other party's behavior combined with the sticker.

At this stage, the MLLM detects and resolves conflicts between $Sen_{is}$ and $Sen_{ts}$ to determine the sticker sentiment $Sen_s$, then resolves conflicts between $Sen_s$ and $Sen_c$ to determine the user's sentiment $\hat{Sen}$. This process is formalized as follows:
\begin{align}
\hat{Sen} &= \Psi_{M}\left( \Psi_{M}\left( Sen_{is}, Sen_{ts} \right), Sen_c \right)
\end{align}

\textbf{Scenario Reconstruction.}
To further improve intent recognition and reasoning reliability, we draw on situated cognition theory from cognitive science~\cite{cakmakci2025situated} and require the MLLM to reconstruct the conversational scenario. Specifically, the MLLM determines who the user is referring to based on the pronouns in the context to initially establish the scenario. Given the predicted sentiment $\hat{Sen}$, latent intent sets $Int^{lat}_c$ and $Int^{lat}_s$, the MLLM first removes latent intents that contradict $\hat{Sen}$ (e.g., \textit{``praise''} is incompatible with a negative sentiment). Since intent is highly context-dependent~\cite{shi2025impact}, the MLLM prioritizes $Int^{lat}_c$ from context for scenario reconstruction. Finally, the MLLM supplements any previously overlooked latent intents based on the reconstructed scenario and filters for those consistent with the scenario, yielding the candidate latent intent set $Int^{lat}_{can}$.

\textbf{Confusing Intent Discrimination.}
In the final stage, the MLLM leverages the accumulated candidate latent intent set 
$Int^{lat}_{can}$ and confusing intent discrimination rules derived from VRLCID to determine the intent. The MLLM first maps $Int^{lat}_{can}$ to predefined intent labels and obtains the candidate intent set $Int_{can}$, and then examines whether any intent labels have been omitted based on the reconstructed conversational scenario. For example, given the context \textit{``This is my friend John, he is excellent at playing soccer.''}, the reconstructed event indicates that the user is introducing a friend and praising their skill. This corresponds to the intent labels \textit{``introduce''} and \textit{``praise''}, and any missing label is added to $Int_{can}$. Even after this supplementation, $Int_{can}$ may still contain semantically similar intent pairs that are difficult to distinguish. For instance, action-oriented intents (such as \textit{``introduce''}) should be prioritized over emotion-oriented ones (such as \textit{``praise''}) in intent consideration, unless the user is explicitly expressing emotions~\cite{hommel2017contributions}. However, most confusing intent pairs exhibit far more subtle distinctions, and pre-trained knowledge alone is insufficient for the MLLM to resolve them reliably. Therefore, we design the VRLCID that enables the MLLM to autonomously derive discrimination rules $\textbf{R}_{\text{conf}}$ and determine the predicted intent $\hat{Int}$:
\begin{align}
\hat{Int} = \Psi_M\left(Int_{\text{can}} \mid \textbf{R}_{\text{conf}} \right)
\end{align}



\subsection{Verbalized Rules Learning module for Confusing Intent Discrimination (VRLCID)}
\label{sec:VRLCID}
To address confusing intent discrimination, we develop a Verbalized Rules Learning module grounded in the paradigm of verbalized learning. Specifically, we treat class discrimination rules as learnable parameters to translate the MLLMs' inherent abstract category discrimination capabilities into concrete and instance-adaptive criteria, and optimize these parameters via the synergistic interplay of three specialized agents: a learner, an optimizer, and a regularizer. We implement this module as VRLCID for confusing intent discrimination. The problem formulation, training objective, and optimization details of VRLCID are presented below.

\textbf{Confusing Intent Discrimination.} Generally, a pair of intent labels ($In_i$, $In_j$) is considered confusing if the MLLM generates interpretable reasoning chains under the same context $T_c$ and sticker $S$. Since both $\hat{Int}$ and the ground truth label $Int^*$ are supported by such reasoning chains, any mismatch $\hat{Int} \neq Int^*$ on a given sample implies that ($In_i$, $In_j$) forms a confusing intent pair.

\textbf{Training Objective.} Our objective is to establish a discrimination rule $r_{In_i,In_j}$ for each confusing intent pair. As displayed in Figure \ref{fig3:flowchart}, the MLLM analyzes training data to derive instance-adaptive discrimination criteria for confusing pairs $(\hat{Int}, Int^*)$, and then establishes the rule set $\textbf{R}_{\text{conf}} = \{r_{In_1,In_2}, \ldots, r_{In_{m-1},In_m}\}$, where $m$ denotes the number of predefined labels. These learnable parameters $\textbf{R}_{\text{conf}}$ are optimized through iterative verbalized interactions among three agents including learner, optimizer, and regularizer. Each agent is implemented by an MLLM equipped with a task-specific prompt.

\textbf{Updating $\textbf{R}_{\text{conf}}$ via Learner, Optimizer, and Regularizer.} 
\textit{Learner.} The learner executes the ``forward pass'' and generates predictions conditioned on current parameters. We define the input as $x=[S,T_c]$, the prompt template as $\theta$, and the verbalized learnable parameters as $\textbf{R}_{\text{conf}}$. The output of the learner is formulated as $\Psi_{M}(x|\theta, \textbf{R}_{\text{conf}})$. To ensure the adaptability of $\textbf{R}_{\text{conf}}$ to SP-CoT, the learner employs SP-CoT prompts (Section \ref{sec:cot}). At iteration $t$, the output is:


\begin{align}
f^{t}_{\text{learner}}(x^t) = \Psi_{M}(x^t|\theta_\text{SP-CoT}, \textbf{R}^{t}_{\text{conf}})
\end{align}
where $\theta_\text{SP-CoT}$ denotes the SP-CoT prompt template, and $x^t$, $\textbf{R}^{t}_{\text{conf}}$ represent the input and parameters at iteration $t$, respectively. The output $f^{t}_{\text{learner}}(x^t)$ comprises both the predicted intent and its accompanying reasoning chain. Prediction errors from the batch are used by the optimizer to update the parameters.

\textit{Optimizer.} The optimizer performs error analysis to identify overlooked or misunderstood fine-grained features or critical latent intents, subsequently formulating refined confusing intent discrimination rules. Following \cite{xiao2025verbalized}, the optimizer adopts the same MLLM as the learner. Let $f_{\text{batch}} = [f^{i}_{\text{learner}}(x^i), \ldots, f^{j}_{\text{learner}}(x^j)]$ denote the collection of erroneous predictions in the batch, where $i, j$ index the error iterations and $x^i, x^j$ represent the corresponding inputs. The inputs and ground truths corresponding to $f_{\text{batch}}$ are $x_\text{batch}$ and $Int^*_\text{batch}$. Conditioned on prompt $\theta_{\text{opt}}$, the optimizer generates updated rules $\textbf{R}^{t}_{\text{opt}}$ as:
\begin{align}
\textbf{R}^{t}_{\text{opt}} = \Psi_{M}(f_{\text{batch}},x_\text{batch}, Int^*_\text{batch}, \textbf{R}^{t}_{\text{conf}} \mid \theta_{\text{opt}})
\end{align}

\textit{Regularizer.} 
The optimizer may produce rules that overfit specific training instances. For example, given a sticker depicting water consumption with the label \textit{``guess''}, the optimizer might spuriously associate \textit{``drinking water''} with \textit{``guess''}. To address this, we introduce a regularizer to eliminate spurious correlations and abstract instance-specific features into generalized semantic representations before parameter updates, preventing the model from overfitting:
\begin{align}
\textbf{R}^{t}_{\text{reg}} = \Psi_{M_r}(\textbf{R}^{t}_{\text{opt}}, \textbf{R}^{t}_{\text{conf}} \mid \theta_{\text{reg}})
\end{align}
where $M_r$ denotes the regularizer model and $\theta_{\text{reg}}$ represents its prompt template. Since directly incorporating rules from different models compromises parameter optimization consistency \cite{xiao2025verbalized}, we design an adaptation prompt $\theta_{\text{adapt}}$ for the learner to assimilate $\textbf{R}^{t}_{\text{reg}}$:
\begin{align}
\textbf{R}^{t+1}_{\text{conf}} = \Psi_{M}(\textbf{R}^{t}_{\text{reg}} \mid \theta_{\text{adapt}})
\end{align}

\section{Experiments}
\label{sec:experiments}
In this section, we evaluate ExCoVer on CSMSA~\cite{ge2022towards} and MSAIRS~\cite{shi2025impact}, addressing the following research questions:

{\bf RQ1}: How does ExCoVer compare with existing MCAS and MLLM baselines on sentiment and intent recognition? (Section~\ref{subsec:comparison})

{\bf RQ2}: How do SP-CoT and VRLCID individually contribute to overall performance? (Section~\ref{subsec:component}, Appendix~\ref{appendix:gain})

{\bf RQ3}: How sensitive is ExCoVer to key reasoning and rule-learning design choices, including SP-CoT stages, the VRLCID regularizer, and training-sample configuration? (Section~\ref{subsec:component}, Appendix~\ref{appendix:ablation}, Appendix~\ref{appendix:stability})

{\bf RQ4}: How well does ExCoVer generalize across input modalities, task settings, and backbone MLLMs? (Section~\ref{subsec:component}, Appendix~\ref{appendix:opensource})

{\bf RQ5}: Are SP-CoT reasoning chains faithful to and interpretable for the final predictions? (Section~\ref{subsec:explainability}, Appendix~\ref{appendix:repro})

We additionally report the evaluation protocol (Appendix~\ref{appendix:protocol}), method design analysis (Appendix~\ref{appendix:design}), computational cost (Appendix~\ref{appendix:comput}), and limitations (Appendix~\ref{appendix:limitations}) in the \textbf{Appendix}.

\begin{table*}[htbp]
\centering
\caption{Comparison on the CSMSA and MSAIRS datasets. M*S and M*I represent the sentiment and intent recognition tasks, respectively. The best and second-best results are highlighted in \textbf{bold} and \underline{underline}, respectively.}
\label{tab:method_performance}
\resizebox{\linewidth}{!}{%
\begin{tabular}{l l c c c c c c c c c c c c}
\toprule
\multirow{2}*{\textbf{Category}} & \multirow{2}*{\textbf{Method}} & \multicolumn{2}{c}{\textbf{Easy-Task}} & \multicolumn{2}{c}{\textbf{Hard-Task1}} & \multicolumn{2}{c}{\textbf{Hard-Task2}} & \multicolumn{2}{c}{\textbf{Hard-Task3}} & \multicolumn{2}{c}{\textbf{M*S}} & \multicolumn{2}{c}{\textbf{M*I}} \\
\cmidrule(lr){3-4} \cmidrule(lr){5-6} \cmidrule(lr){7-8} \cmidrule(lr){9-10} \cmidrule(lr){11-12} \cmidrule(lr){13-14}
& & \textbf{Acc.} & \textbf{F1} & \textbf{Acc.} & \textbf{F1} & \textbf{Acc.} & \textbf{F1} & \textbf{Acc.} & \textbf{F1} & \textbf{Acc.} & \textbf{F1} & \textbf{Acc.} & \textbf{F1} \\
\midrule
\multirow{3}*{Text-only} & BERT\cite{devlin2019bert} & 61.78 & 57.64 & 52.74 & 47.09 & 54.67 & 51.02 & 51.14 & 36.67 & 65.08 & 64.48 & 61.94 & 61.84 \\
& RoBERTa\cite{liu2019roberta} & 61.20 & 59.41 & 54.78 & 49.82 & 53.13 & 52.16 & 54.31 & 41.94 & 66.78 & 66.55 & 65.74 & 65.40 \\
& DeepSeek-V3.2\cite{liu2025deepseek} &51.59 &53.56 &51.97 &56.11 &44.62 &46.61 &43.12 &44.36 &63.14 &63.85 &55.47 &53.29  \\
\midrule
Image-only & ResNet34\cite{he2016deep} & 61.78 & 57.01 & 57.53 & 55.30 & 53.42 & 47.78 & 48.86 & 34.84 & 42.56 & 38.97 & 12.80 & 11.03 \\
\midrule
\multirow{3}*{MLLM} 
& DouBao1.6\cite{guo2025seed1} &65.28 &65.46 &67.72 &68.16 &63.56 &63.07 &64.39 &65.02 &66.06 &63.79 &54.01 &55.17 \\
& Qwen3-VL-Plus\cite{yang2025qwen3} &65.49 &65.08 &67.48 &67.31 &65.12 &65.01 &64.22 &64.88 &65.20 &63.07 &50.19 &48.24 \\
& Gemini3-Flash\cite{laurent2026google} &69.42 &67.93 &68.50 &67.96 &66.15 &61.24 &66.06 &66.26 &70.07 &67.23 &61.31 &60.12 \\
\midrule
\multirow{4}*{Related domain}
& BEAR\cite{yang2025uncertain} &59.55 &61.09 &45.57 &47.27 &53.84 &53.33 &57.38 &58.71 & 62.46 & 62.18 & 61.81 & 62.28 \\
& LGSRR\cite{zhou2025llm} &- &- &-  &- &- &-  &- &- &- &- & 63.02 &62.16 \\
& EMOE\cite{fang2025emoe} &58.92 &59.12 &57.53 &56.97 &51.33 &53.86 &54.55 &54.71 &64.23 &63.91  &- &-  \\
& CoCoT\cite{park2025cognitive} &65.48 &66.37 &69.28 &68.93 &63.85 &63.99 &65.09 &66.15 &69.56 &67.94 &60.58 &60.15 \\ 
\midrule
\multirow{5}*{Multimodal} & mBERT\cite{devlin2019bert} & 59.24 & 55.76 & 57.53 & 56.56 & 56.00 & 53.61 & 51.31 & 40.37 & 68.86 & 68.95 & 65.50 & 58.40 \\
& SAMSAM\cite{ge2022towards} & 63.69 & 61.80 & 59.59 & 56.56 & 55.33 & 51.79 & 54.55 & 42.65 &- &- &- &- \\
& MMSAIR\cite{shi2025impact} &- &- &-  &- &- &-  &- &- &70.58 &70.91 &72.31 &72.29 \\
& MSA-ITEI\cite{shi2025msa} & \underline{71.66} & \underline{70.67} & \underline{68.84} & \underline{68.69} & \underline{68.67} & \underline{70.44} & \underline{75.00} & \underline{74.09} & \underline{72.54} & \underline{72.52} & \underline{74.31} & \underline{74.21} \\
& \textbf{ExCoVer (Ours)} & \textbf{73.24} & \textbf{73.60} & \textbf{72.44} & \textbf{72.40} & \textbf{71.53} & \textbf{70.92} & \textbf{75.23} & \textbf{76.15} & \textbf{74.09} & \textbf{73.56} & \textbf{75.55} & \textbf{75.43} \\

\bottomrule
\end{tabular}%
}
\end{table*}

\subsection{Experimental Setups}
\textbf{Datasets.} CSMSA and MSAIRS datasets feature conversational contexts paired with stickers. MSAIRS dataset comprises 3,118 context-sticker pairs annotated with 3 sentiment labels (\textit{neutral}, \textit{positive}, \textit{negative}) and 20 intent labels (e.g., \textit{complain}, \textit{thank}, \textit{query}). CSMSA contains 1,564 context-sticker pairs with 3 sentiment labels, partitioned into four subsets based on task difficulty and context-sticker alignment: Easy-Task, Hard-Task1, Hard-Task2, and Hard-Task3. As CSMSA does not contain intent annotations, multimodal intent recognition is evaluated only on MSAIRS, whereas sentiment analysis is performed on both datasets. Following standard data splits, ExCoVer is trained on the MSAIRS training set and evaluated on the test sets of both CSMSA and MSAIRS.

 \textbf{Evaluation Metrics.}
Following the evaluation protocol of \cite{shi2025impact}, we report both Accuracy (Acc.) and weighted F1-score (F1) for sentiment and intent classification. These metrics quantify the agreement between predicted labels and ground truth annotations.

 \textbf{Implementation Details.}
The $\Psi_M$ of our method is DouBao-Seed-1.6-Vision (abbreviated as DouBao1.6)~\cite{guo2025seed1}. In VRLCID, the same model instantiates both the learner and optimizer agents, while the regularizer utilizes Gemini-3-Flash-Preview (abbreviated as Gemini3-Flash)~\cite{laurent2026google} to leverage broader pre-trained knowledge. VRLCID operates with a batch size of 4. To evaluate the $\textbf{R}_\text{conf}$, we have a validation set which is 10\% of samples randomly drawn from the original training set of MSAIRS and the remaining data is used as the training set. Following empirical analysis detailed in Section~\ref{subsec:component}, we adopt balanced sampling to construct a training set of 200 samples. During both training and inference, all interactions with LLMs and MLLMs are conducted via API calls with temperature fixed at 0 and all other parameters set to their defaults. All experiments use the above configuration and are run on a server with two NVIDIA V100 GPUs (32 GB memory per GPU).

\subsection{Comparison Results}
\label{subsec:comparison}
We compare ExCoVer against various unimodal and multimodal methods. Text-only models include BERT~\cite{devlin2019bert}, RoBERTa~\cite{liu2019roberta}, and DeepSeek-V3.2~\cite{liu2025deepseek}; the image-only model is ResNet34~\cite{he2016deep}. Multimodal baselines include mBERT~\cite{devlin2019bert}, MMSAIR~\cite{shi2025impact}, SAMSAM~\cite{ge2022towards}, and MSA-ITEI~\cite{shi2025msa}. 
For MLLMs, we evaluate DouBao1.6, Qwen3-VL-Plus~\cite{yang2025qwen3}, and Gemini3-Flash. To comprehensively validate our approach, we additionally compare against methods from related domains: the joint learning method BEAR~\cite{yang2025uncertain}, the multimodal intent recognition method LGSRR~\cite{zhou2025llm}, the multimodal emotion recognition method EMOE~\cite{fang2025emoe}, and the general-purpose CoT method CoCoT~\cite{park2025cognitive}.

Table \ref{tab:method_performance} presents the experimental results across all models. ExCoVer attains 75.55\% accuracy on MSAIRS intent recognition, improving upon MMSAIR, MSA-ITEI, and DouBao1.6 by 3.24\%, 1.24\%, and 21.54\%, respectively. Although MSA-ITEI also employs MLLMs, it only uses them to generate the caption of the sticker for auxiliary reasoning. In contrast, ExCoVer explicitly models sentiment-intent alignment and learns discriminative rules for ambiguous intents through VRLCID. As a result, ExCoVer achieves state-of-the-art performance on both CSMSA and MSAIRS while providing explicit reasoning chains.



Text-only results show that BERT and RoBERTa outperform DeepSeek-V3.2 due to differences in training objectives. Encoder models like BERT excel at intent classification because their masked pre-training enables bidirectional context modeling for semantic-label mappings, whereas generative LLMs like DeepSeek-V3.2 prioritize next-token prediction over precise discrimination~\cite{qorib2024decoder}. Similarly, mBERT outperforms MLLMs on intent recognition. For sentiment classification, MLLMs like DouBao1.6 achieve competitive performance by leveraging scale-driven emotional knowledge~\cite{zhang2024sentiment} and instruction-tuning capabilities~\cite{zhang2026instruction} to compensate for weaker discriminative capacity. Image-only methods perform reasonably on sentiment but poorly on intent, which confirms that intent information primarily resides in conversational context. Among MLLMs, Gemini3-Flash outperforms DouBao1.6 and Qwen3-VL-Plus due to its superior reasoning abilities from pre-training. While CoCoT achieves moderate results, it underperforms ExCoVer which is specifically designed for MCAS. Methods excelling in multimodal intent or emotion recognition, such as BEAR, perform poorly on sticker-context multimodal analysis. This performance gap stems from a fundamental modality mismatch: traditional multimodal intent recognition methods are designed for video modalities, employing temporal dependency modeling to achieve cross-modal synchronization (vision-audio-text) for capturing event-driven intent in video content. In contrast, MCAS involves static modalities (sticker images and chat context), requiring semantic-level intent alignment across modalities rather than temporal alignment~\cite{shi2025impact}. This fundamental difference renders the core modules of traditional methods ineffective in MCAS, leading to significant transfer learning bottlenecks.

\subsection{Component Analysis}
\label{subsec:component}
To validate ExCoVer's design, we conduct component ablations, single-task and modality analyses, backbone transfer experiments, and VRLCID configuration studies on MSAIRS.

\begin{table}[t]
  \centering
  \caption{Ablation study of each component of ExCoVer, modality, single-
  task and different MLLMs on ExCoVer.}
  \label{tab:ablation-excover} 
  \small 
  \begin{tabular*}{\columnwidth}{@{\extracolsep{\fill}} lcccc}
    \toprule
    \multirow{2}{*}{\textbf{Method}} & \multicolumn{2}{c}{\textbf{M*S}} & \multicolumn{2}{c}{\textbf{M*I}} \\
    \cmidrule(lr){2-3} \cmidrule(lr){4-5} 
    & \textbf{Acc.} & \textbf{F1} & \textbf{Acc.} & \textbf{F1} \\
    \midrule
    \multicolumn{5}{@{}l}{\textbf{Each Component of ExCoVer}} \\
     w/o SP-CoT & 66.72 & 64.02 & 55.84 & 56.78 \\
     w/o VRLCID  & 73.58 & 72.49 & 70.51 & 69.58 \\
     \midrule
    \multicolumn{5}{@{}l}{\textbf{Single-Task}} \\
    Sentiment Recognition & 74.23 & 72.92 & - & - \\
    Intent Recognition& - & - & 67.23 & 68.09 \\
    \midrule
    \multicolumn{5}{@{}l}{\textbf{Unimodal}} \\
    Only Context & 62.72 & 64.06 & 63.49 & 62.83  \\
    Only Sticker & 59.12 & 58.09 & 29.20 & 28.35  \\
    \midrule
    \multicolumn{5}{@{}l}{\textbf{Different MLLMs on ExCoVer}} \\
    Qwen3-VL-Plus &73.21  &72.33 &72.85  &71.28    \\ 
    DeepSeek-V3.2 &63.86  &63.25 &64.59 &64.33    \\
    \bottomrule
  \end{tabular*}
\end{table}

\begin{table}[!]
  \centering
  \caption{Ablation study of each stage on SP-CoT and regularizer variants on VRLCID.}
  \label{tab:ablation}
  \small
  \begin{tabular*}{\columnwidth}{@{\extracolsep{\fill}}lcccc}
    \toprule
    \multirow{2}{*}{\textbf{Method}} & \multicolumn{2}{c}{\textbf{M*S}} & \multicolumn{2}{c}{\textbf{M*I}} \\
    \cmidrule(lr){2-3} \cmidrule(lr){4-5}
    & \textbf{Acc.} & \textbf{F1} & \textbf{Acc.} & \textbf{F1} \\
    \midrule
    Full method &74.09 &73.56 &75.55 &75.43  \\
    \midrule
    \multicolumn{5}{@{}l}{\textbf{Key Stage on SP-CoT}} \\
    w/o Stage1 & 70.58 & 69.16 & 66.28 & 66.03 \\
    w/o Stage2  & 67.15 & 66.52 & 68.03 & 67.84 \\
    w/o Stage3  & 72.63 & 72.91 & 69.42 & 68.25 \\
    w/o Stage4 & 73.28 & 72.39 & 66.86 & 66.73 \\
    \midrule
    \multicolumn{5}{@{}l}{\textbf{Regularizer Replacement on VRLCID}} \\
    w/o Regularizer & 72.77 & 71.90 & 67.23 & 66.29 \\
    Regularizer-DouBao1.6 &72.63  &72.04 &72.99  &73.47    \\
    Regularizer-DeepSeek-V3.2 &73.87  &72.94 &75.04 &74.38    \\

    \bottomrule
  \end{tabular*}
\end{table}

\begin{table}[!t]
\caption{Reasoning faithfulness verification of SP-CoT. Stage $k$ denotes truncation at the $k$-th stage (Stage 4 is the full method).}
\label{tab:faithfulness}
\begin{tabular*}{\columnwidth}{@{\extracolsep{\fill}}lcccccc}
\toprule
\textbf{Metric} & \textbf{Stage4} & \textbf{Stage3} & \textbf{Stage2} & \textbf{Stage1} & \textbf{Err-S} & \textbf{Err-R} \\
\midrule
Agree $\downarrow$ & 100.00 & 88.56 & 75.84 & 68.12 & 47.83 & 81.64\\
$\Delta$Acc $\uparrow$ & 0.00 & 11.44 & 24.16 & 31.88 & 52.17 & 18.36 \\
\bottomrule
\end{tabular*}
\end{table}

\textbf{Component Ablation.}
Table~\ref{tab:ablation-excover} shows that removing SP-CoT causes the largest drop, as it supplies structured semantic inputs for VRLCID; without it, VRLCID operates over undifferentiated candidates and its discrimination degrades substantially. Removing VRLCID also consistently hurts intent recognition, indicating that CoT and MLLMs' pre-trained discrimination alone cannot resolve confusing intents, and that $\textbf{R}_{\text{conf}}$ provides indispensable knowledge. Together, SP-CoT and VRLCID are mutually reinforcing.

\textbf{Single-Task, Modality, and Backbone Transfer.}
Sentiment-only reasoning slightly outperforms the full method on sentiment, whereas intent-only reasoning performs much worse on intent, confirming the benefit of sentiment-prior guidance. Sticker-only inputs yield only 29.20\% intent accuracy, while context-only inputs reach 63.49\%, showing that intent cues primarily reside in conversational context. Replacing $\Psi_M$ with Qwen3-VL-Plus and DeepSeek-V3.2 still yields 22.66\% and 9.12\% intent gains over their vanilla baselines, respectively, indicating that ExCoVer transfers across diverse MLLMs and LLMs.

\begin{table}[t]
\centering
\caption{Performance comparison across different sampling strategies on the validation set. The best result is highlighted in \textbf{bold}.}
\resizebox{\linewidth}{!}{
\begin{tabular*}{\columnwidth}{@{\extracolsep{\fill}}lcccc}
\toprule
    \multirow{2}{*}{\textbf{Method}} & \multicolumn{2}{c}{\textbf{M*S}} & \multicolumn{2}{c}{\textbf{M*I}} \\
    \cmidrule(lr){2-3} \cmidrule(lr){4-5}
    & \textbf{Acc.} & \textbf{F1} & \textbf{Acc.} & \textbf{F1} \\
\midrule
\textbf{Balanced sampling} &\textbf{75.18} &\textbf{74.64} &\textbf{76.09} &\textbf{74.95} \\
Stratified sampling &74.38  &74.26  &73.94  &72.92  \\
Random sampling &74.41 &74.63  &74.45  &73.74 \\
\bottomrule
\end{tabular*}
}
\label{tab:sample}
\end{table}

\begin{figure}[t]
    \centering
    \includegraphics[width=0.9\linewidth]{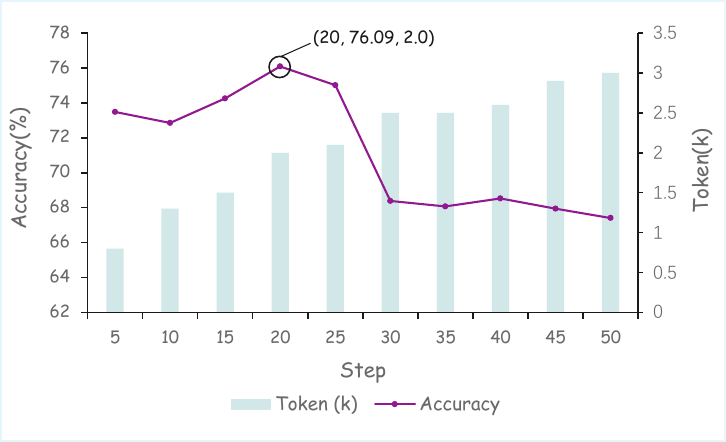}
    \caption{Intent recognition accuracy and token count across 50 training iterations with 200 samples on the validation set.}
    \label{fig:sample}
\end{figure}
\begin{figure*}[t] 
    \centering 
    \includegraphics[width=0.92\textwidth]{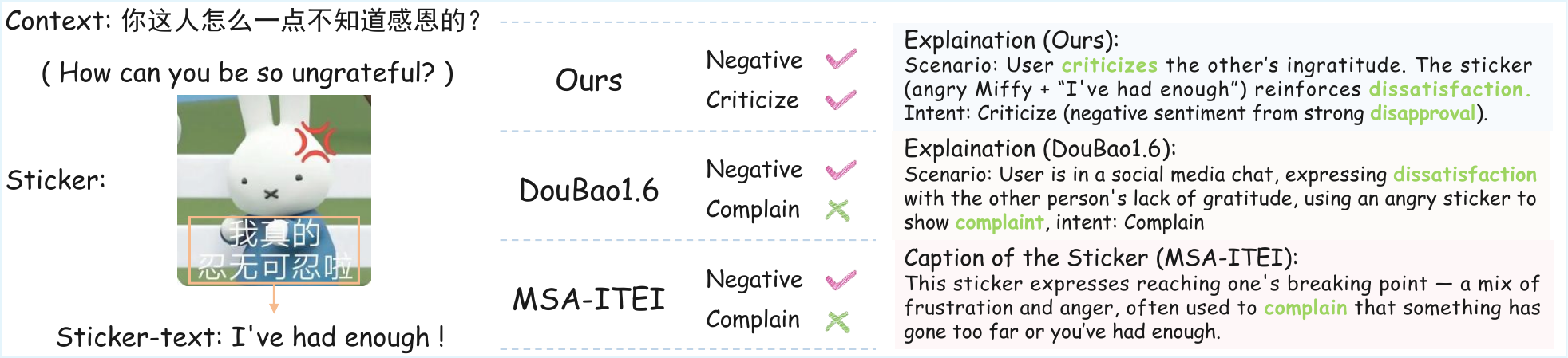} 
    \caption{The visualization of the outputs from ExCoVer, DouBao1.6 and MSA-ITEI
    along with their generated content. The reasoning chain of ours captures core elements such as \textit{``criticizes''}.}
    \label{fig:case} 
\end{figure*}

 \textbf{SP-CoT.} 
Table \ref{tab:ablation} reports ablation results for SP-CoT components. Stage1, Stage2, Stage3, and Stage4 represent Fine-grained Guidance Reasoning, Sentiment Conflict Detection, Scenario Reconstruction, and Confusing Intent Discrimination. We follow the evaluation methodology of \cite{tang2025reason} to assess the contribution of each intermediate reasoning stage through quantifying performance degradation after individual component removal. Ablating any stage consistently degrades performance, which confirms that each reasoning stage contributes significantly to the predictions. In particular, removing Stage1 or Stage4 leads to relatively larger performance drops, highlighting the importance of Fine-grained Guidance Reasoning and Confusing Intent Discrimination to the overall effectiveness of SP-CoT. Removing scenario reconstruction (Stage3) appears to cause minimal impact on sentiment reasoning, as the model implicitly constructs a partial scenario understanding while handling sentiment conflicts in Stage2.

 \textbf{VRLCID.} 
Table \ref{tab:ablation} shows the ablation results of VRLCID. In regularizer replacement experiments, removing the regularizer entirely results in an 8.32\% drop in intent accuracy, and a 3.28\% drop relative to the variant without $\textbf{R}_\text{conf}$. This demonstrates that rules learned solely via the optimizer show limited effectiveness on test sets and sometimes even disrupt inference. This occurs because these rules inherit training set biases without proper filtering. Using the same model as both optimizer and regularizer produces similar issues. The intent accuracy remains significantly lower than the full method. This happens because the optimizer embeds its inherent biases into the learned rules, and the regularizer cannot correct these biases when it shares the same model architecture. In contrast, using a different model (e.g., DeepSeek-V3.2) as the regularizer substantially improves performance approaching the full method, which highlights the importance of using a regularizer distinct from the learner. We omit the individual ablation of learner and optimizer agents since their removal disables the update of $\textbf{R}_\text{conf}$ and leads to complete failure of VRLCID with identical performance to the direct ablation of the VRLCID.

\textbf{Sample Size and Sampling Strategy for VRLCID.}
Since MLLM performance is sensitive to context length \cite{xiao2025verbalized, liu2024lost}, we determine the appropriate size of $\textbf{R}_{\text{conf}}$ for VRLCID by monitoring accuracy against token count (Fig. \ref{fig:sample}). Accuracy peaks when $\textbf{R}_{\text{conf}}$ reaches $\sim$2.0k tokens (at 80 samples); further growth causes overgeneralization. Additionally, we compare balanced, stratified, and random sampling (Table \ref{tab:sample}). Balanced sampling performs best, suggesting that minority classes provide critical discriminative rules. Stratified sampling exhibits notably weaker intent recognition despite preserving the original class distribution, suggesting that minority classes contribute disproportionately critical discrimination rules to $\textbf{R}_{\text{conf}}$. Consequently, ExCoVer adopts balanced sampling with 200 samples as the standard configuration, which makes $\textbf{R}_{\text{conf}}$ reach 2.0k tokens as closely as possible, selecting the best-performing $\textbf{R}_{\text{conf}}$ from the validation set for final evaluation.

\subsection{Explainability Analysis}
\label{subsec:explainability}
\textbf{Reasoning Faithfulness.}
Following \cite{lanham2023measuring}, we conduct truncation and mistake injection experiments on samples correctly predicted by the full method to quantify the faithfulness of SP-CoT reasoning. We report Agree (the ratio of original predictions maintained after intervention) and $\Delta$Acc (the drop in accuracy relative to the full method) to quantify causal reliance on reasoning stages.
As shown in Table \ref{tab:faithfulness}, earlier truncation leads to a progressive decline in Agree, with Stage1 causing a 31.88\% drop, confirming the necessity of the full reasoning chain. For mistake injection, we flip intermediate outcomes: Err-S inverts the result of sentiment detection, and Err-R swaps classification labels within $\textbf{R}_{\text{conf}}$. Err-S causes a 52.17\% performance collapse, highlighting that sentiment-based prior reasoning is pivotal to the final decision. Err-R yields a $\Delta$Acc (18.36\%) higher than that of truncation to Stage3 (11.44\%), indicating that the model does not ignore the corrupted rules but actively follows them, which even impairs its ability to resolve previously solvable intents. These results suggest that SP-CoT's predictions are largely driven by its internal reasoning steps. The high sensitivity to intermediate interventions indicates that post-hoc rationalization is not the dominant factor, thereby validating the faithfulness of our method.

\textbf{Case Study.}
Figure \ref{fig:case} shows two representative examples from MSAIRS with predictions from ExCoVer, DouBao1.6, and MSA-ITEI. The case shows criticism where the user expresses anger at the other party's ingratitude. All three models correctly identify negative sentiment, but DouBao1.6 and MSA-ITEI predict only \textit{``complain''} as the intent, demonstrating that confusing intents pose challenges even for joint learning methods. Examination of the reasoning chains and sticker captions generated by the MLLMs reveals that both DouBao1.6 and MSA-ITEI only identify \textit{``complain''} as the intent. In contrast, ExCoVer's explicit reasoning chain shows it accurately identifies \textit{``criticize''} as the primary intent behind the dissatisfaction by applying discrimination rules learned through VRLCID to distinguish these categories. This result demonstrates that VRLCID effectively supports fine-grained distinction between semantically similar intents.


\section{Conclusion}
\label{sec:conclusion}
We propose ExCoVer, a framework for sentiment and intent recognition in multimodal chat analysis of social media stickers. We extend traditional joint learning by exploring task complementarity throughout the reasoning process. Specifically, our SP-CoT module leverages sentiment as a high-level affective prior to guide intent recognition. This sentiment-prior reasoning explicitly exploits the intrinsic correlation between sentiment and intent, effectively mitigating adverse inter-task interference while providing a narrowed, high-quality intent space for downstream discrimination. Furthermore, we introduce VRLCID, which formulates class discrimination rules as learnable parameters to ground the model's inherent abstract discriminative capabilities into instance-adaptive criteria. Through the collaborative workflow of the learner, optimizer, and regularizer agents, VRLCID systematically refines these rules for confusing intent discrimination, and suppresses spurious correlations to decouple essential intent semantics from non-essential observational noise. Extensive experiments on CSMSA and MSAIRS benchmarks demonstrate that ExCoVer achieves state-of-the-art performance on both sentiment and intent recognition while simultaneously providing human-readable reasoning chains for each prediction. 









\bibliographystyle{ACM-Reference-Format}
\bibliography{main}

\appendix

\clearpage
\twocolumn[ 
\centering
\huge\bfseries Appendix\\
\vspace{1cm} 
]   
\definecolor{promptbg}{RGB}{248, 249, 250}       
\definecolor{promptframe}{RGB}{107, 114, 128}     
\definecolor{promptaccent}{RGB}{156, 163, 175}    
\definecolor{promptaccentsoft}{RGB}{229, 231, 235} 
\definecolor{prompttitlebg}{RGB}{55, 65, 81}      
\definecolor{prompttitletext}{RGB}{243, 244, 246} 
\tcbset{
	promptboxstyle/.style={
		enhanced,
		breakable,
		colback=promptbg,
		colframe=promptframe,
		boxrule=0.4pt,
		arc=1.5pt,
		left=8pt,
		right=7pt,
		top=6pt,
		bottom=6pt,
		borderline west={2.8pt}{0pt}{promptaccent},
		borderline east={0.4pt}{0pt}{promptaccentsoft},
		colbacktitle=prompttitlebg,
		coltitle=prompttitletext,
		fonttitle=\bfseries\sffamily\footnotesize,
		title filled,
		titlerule=0pt,
		toptitle=2.5pt,
		bottomtitle=2.5pt,
		attach boxed title to top left={
			yshift=-2mm,
			xshift=4mm
		},
		boxed title style={
			enhanced,
			colback=prompttitlebg,
			colframe=promptaccent,
			boxrule=0pt,
			arc=1pt,
			top=1.5pt,
			bottom=1.5pt,
			left=5pt,
			right=5pt,
			borderline south={0.8pt}{0pt}{promptaccent},
		},
		underlay unbroken and first={
			\fill[promptaccentsoft, opacity=0.35]
			([xshift=2.8pt]frame.north west) rectangle ([xshift=7pt]frame.south west);
		},
		overlay unbroken and first={
			\draw[promptaccent, line width=0.6pt]
			([xshift=2.8pt, yshift=-4pt]frame.north west) -- ++(0,-6pt);
			\draw[promptaccent, line width=0.6pt]
			([xshift=2.8pt, yshift=4pt]frame.south west) -- ++(0,6pt);
		},
	}
}

\section{Evaluation Protocol and Experimental Setup}
\label{appendix:protocol}

This section clarifies the train/validation/test protocol and the experimental settings referenced in Section~\ref{sec:experiments} of the main manuscript.

\textbf{Validation vs.\ Test Split.}
Table~\ref{tab:sample} and Figure~\ref{fig:sample} in the main text report \textbf{validation set} performance, used for sampling-strategy comparison and $\textbf{R}_{\text{conf}}$ checkpoint selection. Table~\ref{tab:method_performance} reports \textbf{test set} performance using the selected $\textbf{R}_{\text{conf}}$. The workflow is: balanced sampling produces candidate $\textbf{R}_{\text{conf}}$ $\rightarrow$ the best checkpoint is selected on the validation set $\rightarrow$ the selected rules are evaluated on the held-out test set. The numerical difference between validation and test results (e.g., 76.09\% vs.\ 75.55\% intent accuracy under balanced sampling) arises naturally from this split. The 10\% validation set described in Section~\ref{sec:experiments} is reserved for \textbf{selecting the best} $\textbf{R}_{\text{conf}}$ checkpoint; the 200 balanced samples define the \textbf{training} subset for $\textbf{R}_{\text{conf}}$ optimization---these are separate splits serving different purposes. Although accuracy peaks at ${\sim}80$ samples (${\sim}2.0$k tokens) in Figure~\ref{fig:sample}, we adopt 200 balanced samples to ensure all intent classes are sufficiently represented, guaranteeing that the rule set reliably reaches the ${\sim}2.0$k-token effective capacity regardless of sampling randomness.

\textbf{Train/Test Protocol and CSMSA Comparability.}
SP-CoT is a few-shot prompting pipeline with two synthetic examples, requiring no training data. VRLCID's $\textbf{R}_{\text{conf}}$ is learned exclusively from the MSAIRS training set (CSMSA lacks intent labels). CSMSA results therefore reflect SP-CoT's \textbf{few-shot} sentiment reasoning without any CSMSA training data. ExCoVer is thus at a strict disadvantage compared to baselines trained on CSMSA's own split, yet still achieves competitive or superior performance on sentiment recognition.

\textbf{Ablation Definitions.}
``w/o SP-CoT'' replaces the four-stage SP-CoT pipeline with direct prompting while retaining $\textbf{R}_{\text{conf}}$; ``w/o VRLCID'' uses the full SP-CoT pipeline without learned discrimination rules. The coupling variant (see Table~\ref{tab:add_ablation}) retrospectively revises sentiment after predicting intent, testing whether intent should guide sentiment rather than vice versa.

\textbf{DeepSeek-V3.2 Text-Only Variant.}
When DeepSeek-V3.2 serves as $\Psi_M$ in Table~\ref{tab:ablation-excover}, the sticker image is \textbf{omitted entirely}: DeepSeek operates on context ($T_c$) and sticker text ($T_s$) only, with no caption or OCR proxy. This variant tests ExCoVer's \textbf{adaptability to text-only LLMs}; all SP-CoT stages except cross-modal sentiment conflict detection are modality-agnostic and directly applicable. The resulting +9.12\% intent accuracy gain confirms framework transferability rather than serving as a direct multimodal comparison against vision-capable MLLMs.

\section{Additional Experiments and Results}
\label{appendix:ablation}

To provide deeper insights into the contribution of each component in our proposed framework, additional ablation studies are conducted for VRLCID and SP-CoT. Table~\ref{tab:add_ablation} presents quantitative results across different ablation configurations on the MSAIRS dataset.

\begin{table}[t]
	\centering
	\caption{Additional ablation study results on MSAIRS dataset. We evaluate the contribution of key components including the adaptation prompt, few-shot examples, and latent intent module. The coupling variant of ExCoVer evaluates the sentiment after inferring the intent, which reflects the bidirectional interplay between the two tasks.}
	\label{tab:add_ablation}
	\resizebox{\linewidth}{!}{
		\begin{tabular}{l@{\hspace{0.8em}}c@{\hspace{0.6em}}c@{\hspace{0.8em}}c@{\hspace{0.6em}}c}
			\toprule
			\multirow{2}{*}{\textbf{Method}} & \multicolumn{2}{c}{\textbf{M*S}} & \multicolumn{2}{c}{\textbf{M*I}} \\
			\cmidrule(lr){2-3} \cmidrule(lr){4-5}
			& \textbf{Acc.} & \textbf{F1} & \textbf{Acc.} & \textbf{F1} \\
			\midrule
			Full method & 74.09 & 73.56 & 75.55 & 75.43 \\
			w/o $\textbf{R}_\text{conf}$ & 73.58 & 72.49 & 70.51 & 69.58 \\
			\midrule
			\multicolumn{5}{l}{\textbf{Variants of VRLCID}} \\
			$\textbf{R}_\text{conf}$ generated by the MLLM & 73.80 & 72.68 & 71.02 & 69.27 \\
			w/o $\theta_\text{adapt}$ & 73.46 & 72.65 & 70.94 & 70.29 \\
			\midrule
			\multicolumn{5}{l}{\textbf{Key Components of SP-CoT}} \\
			w/o few-shot examples & 72.26 & 71.30 & 69.34 & 68.33 \\
			w/o latent intent module & 71.39 & 69.71 & 65.84 & 65.43 \\
			\midrule
			\multicolumn{5}{l}{\textbf{Coupling Variant of ExCoVer}} \\
			Coupling Variant & 71.50 & 72.21 & 74.18 & 74.52 \\
			\bottomrule
		\end{tabular}
	}
	
\end{table}

\begin{figure}[htbp]
	\centering
	\includegraphics[width=\linewidth]{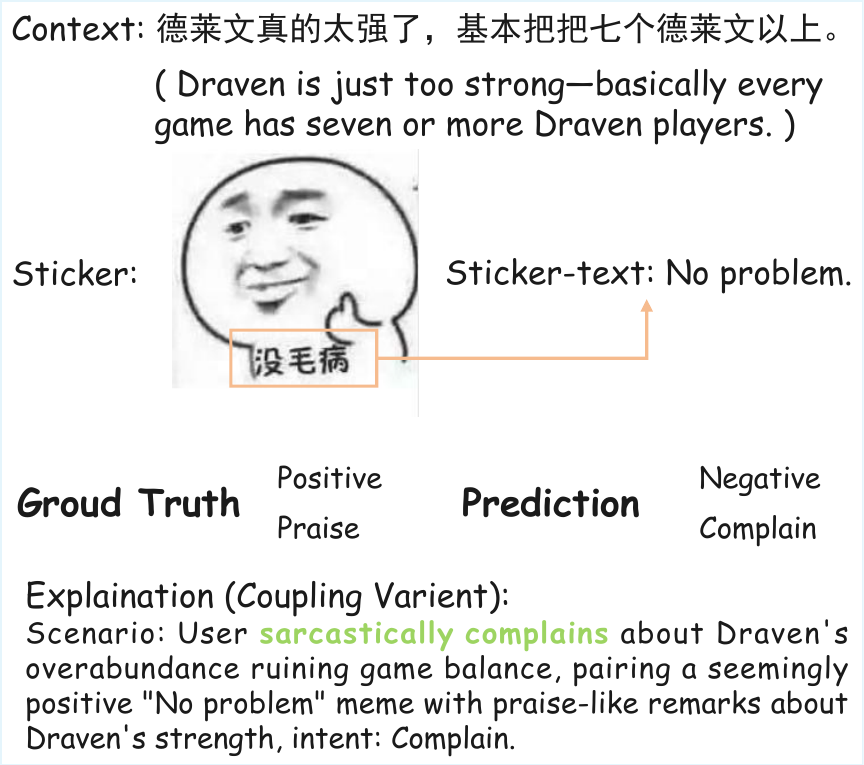}
	\Description{Case study figure: the coupling variant misclassifies praise for a powerful game character as irony.}
	\caption{The coupling variant incorrectly interprets an example of praising a powerful game character as irony.}
	\label{fig:sample2}
\end{figure}

The experimental results demonstrate that $\textbf{R}_\text{conf}$ generated directly by the MLLM achieves only a 0.51\% improvement in intent accuracy compared to the method without $\textbf{R}_\text{conf}$. This indicates that rules self-generated by MLLMs tend to be overly abstract and detached from concrete scenarios, which limits their practical utility in reasoning. The model without the adaptation prompt $\theta_\text{adapt}$ encounters similar issues, as rules summarized by different MLLMs cannot be directly transferred across models. This finding reaffirms the principle of parameter consistency emphasized in \cite{xiao2025verbalized} and underscores the critical role of $\theta_\text{adapt}$. Removing few-shot examples leads to performance degradation in both sentiment and intent recognition, which suggests that these examples effectively help the model comprehend sentiment reversal phenomena and supplement intent based on reconstructed scenarios. The removal of the latent intent module results in more substantial performance drops across both tasks. This occurs because without latent intent, the model forcibly maps predefined intent labels onto each fine-grained token, which essentially imposes rigid associations between these tokens and labels. Such an approach introduces noise inherent in dataset labels and distorts the semantic meanings of tokens, which severely disrupts the reasoning process for both tasks. Finally, regarding the coupling variant of ExCoVer, this variant retrospectively examines alternative sentiment interpretations after predicting intent. While this method partially mitigates cascading errors caused by incorrect sentiment predictions in linear reasoning, it introduces new complications. For instance, in Figure~\ref{fig:sample2}, when a user explicitly praises a game character for being powerful, the retrospective sentiment verification may lead the MLLM to misinterpret the praise intent as sarcasm, where the user is supposedly criticizing the character for being overpowered and disrupting game balance. Due to such errors, the coupling variant exhibits a 1.37\% decrease in intent accuracy and a 2.59\% decrease in sentiment accuracy compared to the full method. These results indicate that employing linear reasoning in ExCoVer constitutes a reasonable and effective strategy for the MCAS task.

\subsection{Performance Gain Decomposition}
\label{appendix:gain}

The 21.54\% intent accuracy gain of ExCoVer over DouBao1.6 direct prompting (75.55\% vs.\ 54.01\% in Table~\ref{tab:method_performance}) decomposes into two complementary contributions: SP-CoT contributes +16.5\% (54.01\%$\rightarrow$70.51\%, corresponding to ``w/o VRLCID'' in Table~\ref{tab:ablation-excover}), and VRLCID contributes +5.04\% (70.51\%$\rightarrow$75.55\%). SP-CoT is fully deterministic under fixed prompts with temperature${=}0$; the majority of gains therefore stem from structured sentiment-prior reasoning rather than rule learning.

\subsection{VRLCID Stability Analysis}
\label{appendix:stability}

The random seed controls VRLCID's 200-sample balanced sampling from the MSAIRS training set. We re-ran VRLCID with four independent random seeds; Table~\ref{tab:stability} reports intent recognition results on the MSAIRS test set. The standard deviation of intent accuracy ($\pm$0.68\%) confirms that VRLCID optimization is stable across random seeds under balanced sampling.

\begin{table}[t]
	\centering
	\caption{VRLCID stability over four random seeds on the MSAIRS test set (intent recognition).}
	\label{tab:stability}
	\small
	\begin{tabular}{lcc}
		\toprule
		\textbf{Seed} & \textbf{Acc. (\%)} & \textbf{F1 (\%)} \\
		\midrule
		1 & 74.67 & 74.16 \\
		2 & 75.84 & 75.71 \\
		3 & 74.82 & 74.51 \\
		4 & 76.28 & 75.94 \\
		\midrule
		Mean & 75.40$\pm$0.68 & 75.08$\pm$0.76 \\
		\bottomrule
	\end{tabular}
\end{table}

\subsection{Open-Source Model Validation}
\label{appendix:opensource}

To verify that ExCoVer is not locked to proprietary APIs, we evaluate a fully open-source combination: Qwen3.5-122B-A10B~\cite{qwen3.5} as the backbone MLLM with DeepSeek-V4-flash~\cite{xu2026deepseek} as the regularizer. Table~\ref{tab:opensource} compares this configuration against the Qwen3.5-122B-A10B direct-prompting baseline on MSAIRS. ExCoVer yields substantial gains on both sentiment and intent recognition, confirming generalization across open-source model families.

\begin{table}[t]
	\centering
	\caption{Open-source model validation on MSAIRS. ExCoVer (open-source) employs Qwen3.5-122B-A10B as the backbone and DeepSeek-V4-flash as the regularizer.}
	\label{tab:opensource}
	\resizebox{\linewidth}{!}{
		\begin{tabular}{lcccc}
			\toprule
			\multirow{2}{*}{\textbf{Method}} & \multicolumn{2}{c}{\textbf{M*S}} & \multicolumn{2}{c}{\textbf{M*I}} \\
			\cmidrule(lr){2-3} \cmidrule(lr){4-5}
			& \textbf{Acc.} & \textbf{F1} & \textbf{Acc.} & \textbf{F1} \\
			\midrule
			Qwen3.5-122B-A10B (baseline) & 66.52 & 65.87 & 52.34 & 50.82 \\
			ExCoVer (open-source) & 73.48 & 72.81 & 73.21 & 71.95 \\
			\bottomrule
		\end{tabular}
	}
\end{table}

\section{Method Design Analysis}
\label{appendix:design}

\subsection{Sentiment-Prior Hypothesis Space Reduction}
\label{appendix:entropy}

\begin{figure}[htbp]
	\centering
	\includegraphics[width=\columnwidth]{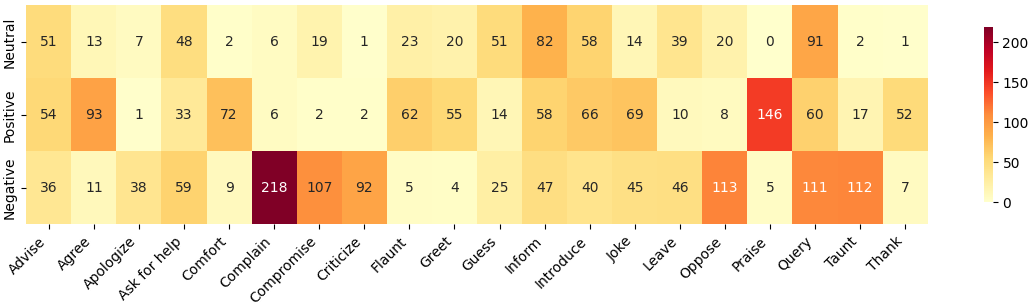}
	\caption{Distributions of intent and sentiment labels in the MSAIRS training set.}
	\label{fig:msairs_label_distribution}
\end{figure}

We quantify the hypothesis-space reduction from sentiment-prior reasoning using conditional entropy on the MSAIRS training set. The label distributions are shown in Figure~\ref{fig:msairs_label_distribution}. The marginal intent entropy is $H(\text{Intent}) = 4.21$ bits, and the conditional entropy given sentiment is $H(\text{Intent} \mid \text{Sentiment}) = 3.73$ bits, yielding an information gain of 0.48 bits (11.42\%). This empirically confirms that sentiment knowledge reduces intent uncertainty and validates the sentiment-prior design of SP-CoT.

Further analysis of sentiment--intent co-occurrence on the training set reveals that sentiment polarity partitions the 20-class intent space into largely non-overlapping subsets: positive sentiment eliminates negative-affect intents (e.g., \textit{complain}, \textit{criticize}, \textit{taunt}), and negative sentiment eliminates positive-affect intents (e.g., \textit{praise}, \textit{greet}, \textit{flaunt}). After this filtering, the effective candidate intent set per sample is reduced from 20 to approximately 8--12 labels, consistent with the scenario-reconstruction stage of SP-CoT.

\subsection{VRLCID Pipeline and Regularizer Mechanism}
\label{appendix:vrlcid}

The VRLCID optimization pipeline operates as follows: (1)~the Learner predicts intent using the current $\textbf{R}_{\text{conf}}$; (2)~misclassified samples are batched ($B{=}4$) and sent to the Optimizer, which proposes discrimination rules in natural language; (3)~the Regularizer (Gemini3-Flash) performs \textbf{subtractive pruning} on rules generated by the Optimizer (DouBao1.6)---it removes spurious correlations rather than generating new rules; (4)~the Learner assimilates the pruned rules via the adaptation prompt $\theta_\text{adapt}$ to update $\textbf{R}_{\text{conf}}$. Same-model regularization yields 72.99\% intent accuracy vs.\ 75.55\% for cross-model regularization (2.56\% gap; see Table~\ref{tab:ablation}), and removing the regularizer entirely causes an 8.32\% intent drop. This validates that cross-model pruning eliminates model-specific biases while preserving generalizable discrimination patterns.

\subsection{Comparison Fairness}
\label{appendix:fairness}

ExCoVer requires no gradient-based training: $\textbf{R}_{\text{conf}}$ is learned via 200-sample verbalized optimization, and the backbone MLLM remains frozen throughout. Comparing multi-stage chain-of-thought reasoning against direct prompting follows standard practice in the community (e.g., Affective-CoT~\cite{huang2025affective}, MM-PEAR-CoT~\cite{li2025multimodal}). The three-agent VRLCID pipeline runs only during \textbf{offline training}; at inference, $\textbf{R}_{\text{conf}}$ is frozen and appended to the SP-CoT prompt. Backbone transfer experiments in Table~\ref{tab:ablation-excover} and regularizer replacement experiments in Table~\ref{tab:ablation} further demonstrate generalizability across model families.

\section{Computational Resources and Efficiency Analysis}
\label{appendix:comput}

While ExCoVer demonstrates strong performance on MCAS tasks, it requires substantially more computational resources than MSA-ITEI. The MLLM of ExCoVer is DouBao1.6~\cite{guo2025seed1}, which employs 230B total parameters with 23B activated parameters. MSA-ITEI uses Yi-6B~\cite{young2024yi} for generating sticker captions, with both total and activated parameters at 6B. To ensure a fair comparison, we further evaluate MSA-ITEI-DouBao1.6 by replacing Yi-6B with DouBao1.6 while keeping the rest of the pipeline unchanged.
Table~\ref{tab:resource} presents a quantitative comparison of model scale and inference efficiency between ExCoVer, MSA-ITEI-Yi-6B, and MSA-ITEI-DouBao1.6.

\begin{table}[h]
	\centering
	\caption{Comparison of computational requirements. End-to-end latency is measured per sample on MSAIRS. Token counts are averaged over the test set.}
	\resizebox{\linewidth}{!}{
		\begin{tabular}{l@{\hspace{0.8em}}c@{\hspace{0.8em}}c@{\hspace{0.8em}}c@{\hspace{0.8em}}c}
			\toprule
			\textbf{Method} & \textbf{Total Params} & \textbf{Activated Params} & \textbf{Latency (s)} & \textbf{Tokens/sample} \\
			\midrule
			DouBao1.6 (direct prompt) & 230B & 23B & 6.4 & 0.35k \\
			MSA-ITEI-Yi-6B & 6B & 6B & 3.8 & -- \\
			MSA-ITEI-DouBao1.6 & 230B & 23B & 4.0 & -- \\
			ExCoVer (SP-CoT + VRLCID) & 230B & 23B & 12.7 & 3.0k \\
			\bottomrule		\end{tabular}
	}
	\label{tab:resource}
\end{table}

As detailed in Table~\ref{tab:resource}, ExCoVer incurs a ${\sim}2\times$ end-to-end latency overhead compared to DouBao1.6 direct prompting (12.7s vs.\ 6.4s per sample) for four-stage SP-CoT reasoning. Despite a ${\sim}9\times$ increase in token consumption (3.0k vs.\ 0.35k tokens per sample), latency scales sub-linearly because API fixed costs (image encoding, model loading) do not grow proportionally with input length. This ${\sim}2\times$ cost yields a 21.54\% intent accuracy gain over direct prompting; VRLCID training is offline and one-time, with $\textbf{R}_{\text{conf}}$ frozen at inference. For target applications such as offline content moderation and batch social media analytics, where accuracy directly determines moderation quality, this trade-off is well justified.

Compared to MSA-ITEI, ExCoVer's full pipeline (including all API round-trips) averages 12.7 seconds per sample under our server configuration, which is 3.3$\times$ and 3.2$\times$ longer than MSA-ITEI-Yi-6B (3.8s) and MSA-ITEI-DouBao1.6 (4.0s), respectively. Since MSA-ITEI only utilizes an MLLM for sticker captioning, this increased overhead is primarily attributed to the multi-step reasoning of SP-CoT.

Despite these efficiency limitations, ExCoVer offers several distinct advantages that justify the additional computational cost in the MCAS task. First, the relatively small parameter count of Yi-6B (6B parameters) limits its capability to perform complex CoT reasoning required for the MCAS task. While MSA-ITEI-DouBao1.6 upgrades to a larger model with 230B total parameters, its caption-only pipeline fails to fully leverage the reasoning capabilities of the larger model, as it restricts the MLLM to merely generating textual descriptions of stickers rather than engaging in multi-step analytical reasoning. In contrast, ExCoVer employs the same large-scale MLLM (DouBao1.6) for comprehensive reasoning tasks, which enables the framework to exploit the model's capacity for complex semantic understanding and logical reasoning. Second, the framework provides flexible modeling of inter-task dependencies through its explicit reasoning structure, which allows the model to effectively exploit positive facilitation between tasks while mitigating adverse constraints. Third, ExCoVer generates explainable reasoning chains for each prediction, which provides researchers and practitioners with transparent insights into the model's decision process. In contrast, both MSA-ITEI variants only generate captions for stickers and lack explainability in their predictions.

\section{Reproducibility and Transparency}
\label{appendix:repro}

\textbf{Model Versions.}
All API-based experiments use the following model identifiers:
DouBao-Seed-1.6-Vision (\nolinkurl{doubao-seed-1-6-vision-250815}) for the learner and optimizer, and Gemini-3-Flash-Preview for the regularizer.

\textbf{VRLCID Inspectability.}
Each VRLCID agent has a well-defined, inspectable role. The Optimizer outputs human-readable natural-language discrimination rules; the Regularizer's deletions are logged during training; changes introduced by the adaptation prompt $\theta_\text{adapt}$ are diffable against the pre-adaptation rule set. 
The entire $\textbf{R}_{\text{conf}}$ (${\sim}2.0$k tokens) is a plain-text artifact that can be version-controlled, audited, and manually edited when specific rules underperform, unlike neural fine-tuning, where failure modes are opaque.

\textbf{Faithfulness Evaluation Scope.}
The truncation and mistake-injection experiments in Table~\ref{tab:faithfulness} are conducted on \textbf{only correctly predicted samples} (75.55\% of the MSAIRS test set). Following~\cite{lanham2023measuring}, this restriction is methodologically necessary: on incorrectly predicted samples, one cannot distinguish unfaithful reasoning from insufficient model capability. Within this valid subset, Err-S and Err-R injection confirm that predictions are genuinely driven by the reasoning chain rather than post-hoc rationalization.

\section{Limitations}
\label{appendix:limitations}
While ExCoVer achieves strong performance on MCAS, several limitations merit discussion. First, VRLCID is constrained by MLLM context windows: as verbalized rules accumulate during training, accuracy declines once $\textbf{R}_{\text{conf}}$ exceeds roughly 2k tokens, which limits scalability to settings with many confusing intent pairs or full-dataset rule induction. More compact or hierarchical rule representations may be needed to address this bottleneck. Second, the multi-stage SP-CoT pipeline incurs higher inference latency than lighter baselines (see Appendix~\ref{appendix:comput}), which may restrict deployment in latency-sensitive or resource-constrained settings. Third, our experiments focus on Chinese sticker-based conversations in CSMSA and MSAIRS; although ExCoVer is language-agnostic in principle (only $\textbf{R}_{\text{conf}}$ and prompts require localization), the lack of English-focused datasets leaves cross-lingual transfer untested. Fourth, ExCoVer relies on the reasoning and multimodal capabilities of the backbone MLLM, and substantially weaker base models may reduce the reliability of both the reasoning chains and the learned verbalized rules. Finally, while ExCoVer produces explicit reasoning chains, user-perceived explainability has not yet been validated through human evaluation; we plan to conduct user studies in downstream dialogue systems as future work.

\end{document}